\documentclass[pra,twocolumn,letterpaper,showkeys,showpacs, superscriptaddress]{revtex4-1}
\usepackage{graphicx,amsmath,amssymb,amsfonts,latexsym,color,dcolumn,bm}
\usepackage{natbib}
\usepackage{gensymb}
\usepackage{braket}
\usepackage[utf8]{inputenc}
\usepackage{geometry}
\usepackage{array}
\usepackage{comment}
\usepackage{subfigure}
\newcommand{\ra}{\rangle}
\newcommand{\la}{\langle}

\newcommand{\bo}{\mathbf}
\begin{document}

\title{The Role of Vacuum Fluctuations and Symmetry in the Hydrogen Atom in Quantum Mechanics and SED}

\author{G. Jordan Maclay}
\email{jordanmaclay@quantumfields.com}
\address{Quantum Fields LLC, St. Charles IL 60174}
\date{June 2019}
\begin{abstract}SED has had success modeling black body radiation, the harmonic oscillator, the Casimir effect, van der Waals forces, diamagnetism, and uniform acceleration of electrodynamic systems using the stochastic zero-point fluctuations of the electromagnetic field with classical mechanics.  However the hydrogen atom, with its 1/r potential remains a critical challenge.  Cole and Zou in 2003 and Nieuwenhuizen and Liska in 2015 found that the SED field prevented the electron orbit from collapsing into the proton but eventually the atom became ionized.  We look at the issues of the H atom and SED from the perspective of symmetry of the quantum mechanical Hamiltonian which is used to obtain the quantum mechanical results, and the Abraham-Lorentz equation, which is a force equation that includes the effects of radiation reaction and is used to obtain the SED simulations. We contrast the physical computed effects of the quantized electromagnetic vacuum flucuations with the role of the real stochastic electromagnetic field.
\end{abstract}




\maketitle
\section{Introduction}

    The hydrogen atom has been the testing ground for
    theoretical atomic physics for over a hundred years. The original quantum mechanics was motivated to model the H atom and explain its spectrum \cite{herz}.  That theory was refined into modern quantum mechanics by Bohr, Heisenberg and Schrodinger, Dirac, and others. Precision measurements of hydrogen energy levels by Willis Lamb in 1947 disagreed with the theory, which stimulated the development of Quantum Electrodynamics (QED)which included the effects of the vacuum fluctuations of the quantized electromagnetic field. The hydrogen atom is the fundamental two-body system and perhaps the most important tool of atomic physics and the challenge is to calculate its properties to the highest accuracy possible. The current QED theory is the most precise of any physical theory: "The study of the hydrogen atom has been at the heart of the development of modern physics...theoretical calculations reach precision up to the 12th decimal place...high resolution laser spectroscopy experiments...reach to the 15th decimal place for the $1S-2S$ transition...The Rydberg constant is known to 6 parts in $10^{12}$ \cite{beyer}.
    
    Stochastic electrodynamics (SED) represents an effort to explain quantum phenomena through classical physics done in the presence of a real stochastic electromagnetic field which is the sum of a portion for $T=0$ with spectral energy density $\rho(\omega)=\hbar \omega^{3}/2\pi^{2} c^3$, identical to that of the virtual zero-point vacuum fluctuations of the quantized electromagnetic field in QED, plus a Planckian spectrum for a finite temperature $T>0$. In this paper, we focus on the $T=0$ component. SED has had success modeling the harmonic oscillator, and other linear systems. However, analytical efforts to address the H atom with SED have not been successful, so researchers have turned to numerical calculations to explore the H atom in SED and determine if SED can model the ground state \cite{zhou and cole},\cite{nie}.  The assertion of SED originally made by Boyer \cite{boyer} is that the energy lost by radiation from the accelerating electron bound to the H nucleus is statistically compensated by the energy gained from the fluctuating vacuum field. In a variation of this work, Puthoff has shown that particularly for circular orbits, if a stable ground state of constant energy is achieved, then the energy provided by the stochastic vacuum field exactly compensates for the energy lost by radiation \cite{hal}. Others have asserted that no thermodynamic stable equilibrium will be achieved \cite{claverie}. 
    SED theory has been developed in several books by de la Pena and his collaborators \cite{de la pena} in which they have formulated an approach that claims to lead to the equations of standard quantum mechanics. 
    In this paper, we will contrast the approach of quantum mechanics with that of SED, discussing the role of vacuum fluctuations in both theories. Emphasis is placed on the symmetry of the H atom since this plays such a pivotal role. For both theories we make some simplifying assumptions:
    non-relativistic mechanics, no spin, infinitely heavy proton. We discuss the most recent SED calculations which have shown stability for a limited number of orbits, but ionization for longer times\cite{zhou and cole} \cite{nie}.   

\section{Role of Vacuum Fluctuations in Quantum Mechanics and SED}

In the nineteen-thirties, quantum theory had predicted zero-point vacuum fluctuations. Oppenheimer had calculated that the vacuum fluctuations gave an infinite and therefore unphysical energy shift to the free electron. The theory also predicted that the hydrogen $2S^{1/2}$ and $2P^{1/2}$ levels were degenerate.   In 1947 Lamb and Retherford announced they had measured a shift of $1058$ MHz at the Shelter Island Conference. After the conference, Hans Bethe, on the train ride to Schenectady, where he was consulting for GE, developed a new approach to include the vacuum fluctuations and QED was born. He, and others after him, applied renormalization methods, originally developed by Kramers and others, to deal successfully with the infinities predicted by prior approaches.  Renormalization asserted that the vacuum fluctuations lead to the observed physical mass and charge of the electron.

In QED the vacuum field is typically expressed as a sum over an infinite number of plane waves with all possible momenta $\hbar\bo{k}$ and directions $\bo{k}/k$ with the restriction that the energy $E_{k}$ in each mode is $\hbar \omega_k/2=\hbar k/2c$. The vector potential is \cite{milonni}
\begin{multline}
    \bo{E}(\bo{r},t)= \sum_{\bo{k},\lambda}\sqrt{\frac{\hbar \omega_k}{2  \epsilon_{0} V}}(A_{\bo{k}\lambda}\cos{(\bo{k}\cdot\bo{r}-\omega_{k}t)}\\-B_{\bo{k}\lambda}\sin{(\bo{k}\cdot\bo{r}-\omega_{k\}t})})\  \bo{e}_{\bo{k},\lambda}   
\end{multline}
where the raising and lowering operators obey the commutation rules
\begin{equation}
    [a_{\bf{k} \lambda},a^{\dagger}_{\bf{k'} \lambda'}] =\delta_{\bf{k}\bf{k'}}\delta_{\lambda\lambda'} 
\end{equation}
and the two polarization vectors ($\lambda=1,2$) are orthogonal to $\bo{k}$ so $\bo{k}\cdot \bo{e}_{\bo{k},\lambda}=0$, and  
\begin{equation}
  \bo{e}_{\bo{k},\lambda}\cdot  \bo{e}_{\bo{k},\lambda'} =\delta_{\lambda \lambda'}.
\end{equation}
The quantization volume V is an artifice to avoid infinite volumes.  In this box normalization $k_x=2\pi n_x/L_x$ , $k_y=2 \pi n_y/L_y$ , and $k_z=2 \pi n_z/L_z$ , with $V=L_x L_y L_z$, and the integers $n_x$, $n_y$, and $n_z$ go from -$\infty$ to $+\infty.$
The electric field is $\bo{E(\bo{r},t)}=-\partial\bo{A}(\bo{r},t)/\partial r$ and $\bo{B}(\bo{r},t)=\nabla \times \bo{A}(\bo{r},t)$.

To summarize the properties of the vacuum field in QED: no real photons are present, only random virtual photons of energy $\hbar\omega_{k}/2$ and momentum $h\bo{k}/2c$, with all possible momenta present consistent with Eq 1. The expectation values of the fields vanish but the variances do not. The fields are isotropic (invariant under rotations), invariant under space-time translations (homogeneous), and under boosts (Lorentz invariant). The energy density spectrum which is proportional to $\omega^3$ is also Lorentz invariant.

In QED, the zero-point quantum fluctuations are responsible for the mass and charge renormalization of particles. In this process one postulates the following: 

 \begin{center}
 bare point electron + vacuum fluctuations +  radiative reaction $\rightarrow$\\ 
 electron with physical mass, charge  and effective size about Compton wavelength.
\end{center} \vspace{2pt}  

A similar process occurs for an atom, in which the atom undergoes allowed virtual (energy conserving) transitions due to the vacuum field.  These transitions can be seen as shifting the average energy of the atom.  This is one picture (approximate) of radiative shifts, such as the Lamb shift\cite{maclay}.  The Lamb shift is a small shift, about 1 part in $10^6$. It has been predicted correctly to about 1 part in $10^9$ by QED. In QED, radiative shifts are generally calculated using Feynmann diagrams, in which the atom is depicted as propagating in time, and it absorbs or emits a virtual photon changing its state correspondingly, then a short time later (consistent with the time-energy uncertainty principle) emits or absorbs the same virtual photon and returns to the initial state. This process arises from the $\bo{p}\cdot \bo{A}$ term in the Hamiltonian and can involve all possible photons, but only photons of the resonant frequency will induce transitions.  This model in a sense describes the interaction of the electron with its own radiation field.  In QED this is equivalent to interacting with the ubiquitous virtual fluctuating vacuum field.  The electron mass renormalization in QED is the state independent shift in energy for a bare electron given by the $\bo{p}\cdot \bo{A}$ and the $\bo{A} \cdot \bo{A}$ terms. 

What makes Equation 1 for the vector potential unique to the quantized electromagnetic field is the presence of the operators $a_{\bo{k}\lambda}$ and $a^{\dagger}_{\bo{k}\lambda}$.  The expression for the vector potential or the $\bo{E}(\bo{r},t)$ field can be recast with real (Hermitian) operators into the form:
\begin{multline}
 \bo{E}(\bo{r},t)= \sum_{\bo{k},\lambda}\sqrt{\frac{\hbar \omega_k}{2  \epsilon_{0} V}}(A_{\bo{k}\lambda}\cos{(\bo{k}\cdot\bo{r}-\omega_{k}t)}\\-B_{\bo{k}\lambda}\sin{(\bo{k}\cdot\bo{r}-\omega_{k\}t})})\  \bo{e}_{\bo{k},\lambda}  
\end{multline}
where 
\begin{equation}
    [A_{\bo{k}\lambda},B_{\bo{k'}\lambda'}]=2\  i\ \delta_{\bo{kk'}\lambda\lambda'}
\end{equation}
and the expectation values with respect to the vacuum are $\la A_{\bo{k}\lambda}\ra =\la B_{\bo{k}\lambda}\ra= 0 $
and the variances are $\la A^2_{\bo{k}\lambda}\ra =\la B^2_{\bo{k}\lambda}\ra= 1$.  

Equation 4 appears precisely the same as the equation often used in SED, with the differences that $A_{\bo{k}\lambda}$ and $B_{\bo{k}\lambda}$ are not operators but independent Gaussian variables with average $0$ and variance $1$.  Also for each term the energy taken as the integral $\int_{V}d^{3}r(\frac{\epsilon_{0}}{2}\bo{E}^2+\frac{1}{2\mu_{o}}\bo{B}^{2}) = \hbar \omega_k$. In some cases a convergence factor for the energy may be used \cite{nie}. The SED field is real and is scaled so that the spectral energy density $\rho(\omega)=\hbar\omega^3/2\pi^{2}c^3$  is identical to that of QED. The vacuum field in SED is also invariant under Lorentz transformations, which includes translations, rotations, and boosts.

Vacuum fluctuations are at the heart of SED.  Because the expression for the vacuum field has an infinite number of plane waves, in SED calculations mathematical simplifications are employed to make calculations tractable.  How the vacuum field is treated mathematically is probably one of the most critical aspects of SED modeling.  A finite sum over discrete $\bo{k}$ with a restricted range of $\omega_k$ is generally employed, which may alter the symmetry properties of the vacuum field. In the SED H atom calculations, the frequencies employed have been within four orders of magnitude of the rotational frequency of the electron in the atom\cite{zhou and cole},\cite{nie}. The physical mass and charge of the electron are used, whereas in QED the higher frequency components of the vacuum fluctuations have the effect of giving a bare particle the physical mass and charge of the electron. These high frequency components are not explicitly considered in the SED calculations. Even with such simplifications, millions or billions of waves are included requiring extensive computing power. In SED, the vacuum fluctuations are the source of behavior characteristic of quantum systems, for example, for the harmonic oscillator. Indeed that is the basic contention of SED, that classical physics plus the vacuum field leads to the correct predictions for atomic systems, as does quantum mechanics:\vspace{2pt}

\begin{center}classical mechanics + radiative reaction + point electron with physical mass + real vacuum fluctuations(SED)\\ $\rightarrow$ stable atom with correct behavior. \end{center}\vspace{2pt}

In summary, the role of the vacuum field in SED is to stabilize the electron orbit by providing the energy lost to radiation from the $d^3\bo{r}/dt^3$ term.  Since the electron follows a classical orbit in SED, the orbit must be constantly changing to mimic wavelike behavior shown in quantum mechanics.  Indeed from Earnshaw's theorem there is no stable static arrangement of charges so fluctuations are to be expected.  One interpretation of the orbital behavior is that when the electron loses energy and moves closer to the nucleus, its orbital frequency increases and it interacts more strongly with the higher frequency, higher energy waves in the vacuum field, causing it to accelerate, gain energy and move to a larger radius orbit.  With the larger radius, the orbital frequency slows down, the electron interacts more strongly with the lower frequency waves, loses energy and moves to a smaller orbit.  Thus the electron oscillates from smaller to larger radii \cite{boyer} \cite{cole v}.   

\section{The Hydrogen atom in Quantum Mechanics and Classical Mechanics with No Radiative Reaction}

 In the quantum mechanical H atom there is no radiation from the electron due to the standing wave nature of the electron orbitals when the atom is in a stationary state. The wavefunctions and energy levels are determined solely by the symmetry of the Coulomb potential. The Hamiltonian is 
 \begin{equation}                                  H=\frac{\bo{p}^2}{2m} - \frac{Ze^2}{r}.             
\end{equation}

We discuss the symmetry of H and its implication in the classical case with no radiation since SED is a classical theory. We also outline the very similar effects of the symmetry in the quantum mechanical system.  Classically, the orbits about the Coulomb potential are stable orbits (assuming no radiation), meaning that a perturbation to the position of the electron in the orbit will lead only to small oscillations. In the classical system there are orbits, in the quantum mechanical system there are wavefunctions. Solutions to Schrodinger's equation $H|\psi\ra = E |\psi\ra$ are stationary energy eigenstates. The rotational symmetry of the atom implies that H is invariant under rotation. Rotations are generated by $e^{i\bo{L}\phi/  \hbar}$ where the angular momentum $\bo{L} = \bo{r}\times\bo{p}$, so $[H, \bo{L}] = 0$.  On the other hand $[H, \bo{L}] = -i\hbar d\bo{L}/dt$ so the angular momentum is conserved. For the classical orbit, the energy E and $\bo{L}$ are also constant, and $\bo{L}$ would be normal to the orbital plane. Symmetry operators of H are conserved in time. There is an additional symmetry operator of H, the Runge-Lenz vector $\bo{A}$:
\begin{equation}
 \bo{A}=\frac{\bo{p}\times\bo{L}-\bo{L}\times\bo{p}}{2a}-\frac{mZe^2\bo{r}}{r}   
\end{equation}
where $a=(-2mE)^{1/2}$, the effective momentum. The Runge-Lenz vector is constant, pointing along what would classically be the principal axis of the orbit, and is perpendicular to the angular momentum $\bo{L}\cdot \bo{A}=0$ (see Fig.1). In the classical H atom $e^{i\bo{A}\rho/ \hbar}$ is a generator of
changes in the eccentricity $\epsilon$ of the orbit in such a way that the energy remains unchanged. In quantum mechanics, the Runge-Lenz vector generates a transformation that takes a state of given principal quantum number n and given L ($L\le n-1$) into a superposition of the $n^2$ degenerate states with the same n.  
\begin{figure}
    \centering
    \includegraphics[scale=0.26]{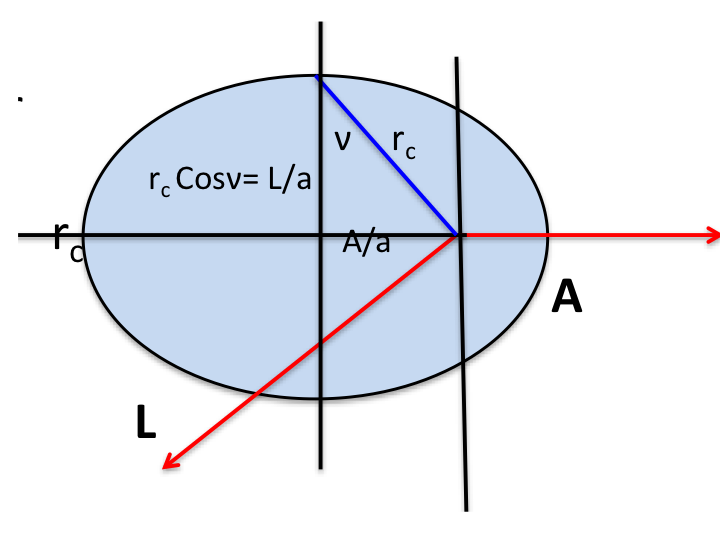}
    \caption{The classical limit of the orbital is a non-precessing ellipse with constant $\bo{L}$ and $\bo{A}$.  The semi-major axis is $r_c$ and $L^2 + A^2 = a^2r^2=\rangle$ energy. The eccentricity $\epsilon=\sin{\nu}$ and $a=(-2mE)^{1/2}$.}
    \label{fig:my_label}
\end{figure}

From the definitions of the two conserved vectors, it is trivial to obtain the equations for the classical orbits:
\begin{equation}
    \bo{r}\cdot\bo{A}=rA\cos{\phi_{r}=-r(mZe^2/a)+\bo{r}\cdot\bo{p}\times\bo{L}},
\end{equation}
and substituting $\bo{r}\cdot\bo{p}\times\bo{L}=L^2$ gives
\begin{equation}
    r=\frac{L^2/mZe^2}{(a/mZe^2)A\cos{\phi_r}+1}
\end{equation}
which describes an ellipse with eccentricity $\epsilon=A(a/mZe^2)$.  Computing $\bo{A}\cdot\bo{A}$ and substituting the value for A gives
\begin{equation}
    \epsilon=[2|E|L^2/m(Ze^2) + 1]^{1/2}.
\end{equation}
The semi-major axis is the average of the radii at the turning points. From the orbit equation, we find
\begin{equation}
    r_c=\frac{L^2}{mZe^2}\frac{1}{1-\epsilon^2}=\frac{Ze^2}{2|E|}.
\end{equation}
The energy depends only on the length of the semi-major axis $r_c$ not on the eccentricity.  This makes sense because $\bo{A}$ changes the eccentricity but not the energy. Using the equations for $A^2$ and $L^2$, with $\epsilon =\sin{\nu}$, we find
\begin{equation}
 L=ar_c\cos{\nu}\hspace{25pt}A=ar_c\sin{\nu}   
\end{equation}
therefore, 
\begin{equation}
 L^2+A^2=a^2r^2_c=-\frac{m(Ze^2)^2}{2|E|}.   
\end{equation}
Thus the values of A and L determine the energy E.
A similar equation can be derived in quantum mechanics, and using the quantization of the operators $\bo{A}$ and $\bo{L}$, we obtain the quantized energy levels as a function of the principal quantum number n.
The commutators of $\bo{A}$ and $\bo{L}$ are
\begin{align}
          [L_i, L_j]&=i\epsilon_{ijk}L_k     
&         [A_i, A_j]&=i\epsilon_{ijk} L_k
&\\         [L_i, A_j]&= i\epsilon_{ijk}A_k.      
\end{align}

The generators ${\bo{A}, \bo{L}}$ form a Lie algebra that closes. To determine the corresponding symmetry group define the generators $\bo{M} =1/2(\bo{L}-\bo{A})$ and $\bo{N}=1/2(\bo{L}+\bo{A})$ which have the commutation properties of two disjoint O(3) groups $O(3) \times O(3) \approx O(4)$:
\begin{align}
[M_i, N_j]&=0     
&[M_i, M_j]=i\epsilon_{ijk} M_k
&\\ [N_i, N_j]&= i\epsilon_{ijk}N_k   .   
\end{align}
The Lie algebra for M or N is the same as the familiar commutation relations for angular momentum $[J_{i}, J_{j}]=i\epsilon_{ijk} J_k$.  For angular momentum the allowed representations are characterized by the scalar $J^2=J(J+1)$, where $J=0, 1/2,1,...$  The $2J+1$ angular moment kets $|Jm\rangle$ for this representation are generally chosen so $J_z|jm\rangle=m|Jm \rangle$, where $m=-J, -J+1,...,0, 1,2,...,J$. In analogy with the usual O(3) results $\bo{M}^2=\bo{N}^2=1/4(\bo{L}^2 +\bo{A}^2)=j(j+1)$ for $j+0, 1/2, 1, ...$ Thus we obtain
\begin{equation}
    \bo{L}^2+\bo{A}^2+1=\frac{(mZe^2)^2}{-2mE}=(2j+1)^2=n^2
\end{equation}
where n=1,2,3,...is the principal quantum number and $n^2$ is the degree of degeneracy for the representation. The wavefunctions are eigenfunction of $L_{z}$ and $A_{z}$. The O(4) symmetry of the wavefunctions is especially manifest in momentum space, where the wavefunctions are spherical harmonics in 4 dimensions. It is possible to enlarge the symmetry group from O(4) to SO(4,2) in such a way that all states, scattering and bound, are included in the representation of the group\cite{maclay}.  

In summary, the symmetry of the Hamiltonian for the H atom determines the energy levels, the orbits or wavefunctions, the degeneracy, the constants of the motion. For the ground state n=1 and L=0 and there is no degeneracy.
\section{The Hydrogen Atom in SED}

In SED, because of the radiative reaction, the energy of the electron is not conserved and the equation of motion generally used is the Abraham-Lorentz force equation\cite{zhou and cole}\cite{cole fop}\cite{nie}
\begin{multline}
    m\frac{d^2\bo{r}}{dt^2}=-\frac{Ze^2\bo{r}}{r^3}+\frac{2e^2}{3c^2}\frac{d^3\bo{r}}{dt^3}\\-e(\bo{E}(\bo{r},t)+\bo{v}\times\bo{B}(\bo{r},t)).
\end{multline}
The first term on the right of the equal sign is the Coulomb force, the next term is the radiative reaction, followed by the Lorentz force due to the vacuum fields $\bo{E}(\bo{r},t)$ and $\bo{B}(\bo{r},t)$.  The presence of the radiative reaction and the Lorentz force eliminates the  symmetry of equation: angular momentum is not conserved, the Runge-Lenz vector is not constant and the electron energy is not constant. It should be mentioned that there is a fully relativistic form of Eq. 19, the Abraham-Lorentz-Dirac equation, used, for example, in the study of charges accelerated uniformly through the vacuum \cite{coleP}.

The Abraham-Lorentz force $(2e^2/3c^2)d\bo{a}/dt$ is the non-relativistic recoil force for an accelerating charged particle caused by the particle emitting radiation which carries momentum, angular momentum and energy\cite{teit}.  The radiation field from the particle is essentially exerting a force on itself, sometimes called a “self-field”, a phenomena which can leads to renormalization and a host of unsolved complications ultimately dealing with the nature of elementary particles\cite{teit}\cite{jackson}.  Radiative reaction and vacuum fluctuations are intimately intertwined\cite{milonni}. For example, in quantum mechanics, both are required to maintain the canonical commutation relations. The third order derivative for the Lorentz force requires the vacuum energy density to go as $\omega^3$.  Renormalization methods in classical physics and in QED have been developed to deal with these issues in an effective if not fundamental way. The Abraham-Lorentz equation is is a third order differential equation in time, which requires three initial conditions (the initial position, velocity and acceleration),  and which leads to mathematical problems such as runaway solutions and acausal behavior and divergences (see \cite{milonniqv} for a discussion).

In most calculations the magnetic field is dropped and a dipole approximation is used so the electric field is a function of time alone $\bo{E}(\bo{r},t)\approx \bo{E}(t)$. The electron is assumed to move in a plane. Also often the third derivative is approximated by using the equation obtained when there is no radiative reaction or vacuum field $md^{2}\bo{r}/dt^{2}=-\frac{Ze^{2}\bo{r}}{r^3}$ so 
$(2e^2/3c^2)d\bo{a}/dt \approx (2Ze^4/3m^2c^2)d\bo{r}/dt$.

\section{SED Simulations of the Ground State of the H Atom}
Cole and Zou performed the first simulations of the hydrogen atom\cite{zhou and cole}\cite{cole v}. They retained the effect of the magnetic field, approximated the radiative term as discussed above, and did not use a dipole approximation. They used an adaptive time step Runge-Kutta 4th order algorithm, roughly equivalent to a net 5th order algorithm in dt to simulate 2 d orbits in a box 27 \AA \ by 27 \AA \ by 0.41 cm, where the normal to the plane of the orbit was in the larger Z direction. The advantage of choosing this geometry was that it reduced the number of plane waves they needed to consider to $2.2\times 10^6$, with wavelengths from 0.1\AA \ to 900\AA \, and waves only in the Z direction, yet was able to capture the most important physical features\cite{zhou and cole}. They did 11 simulations, each starting at a radius of 0.53 \AA \, but with different random amplitudes for the waves.  They ran the simulations for about $10^{-11}$ sec or roughly 100,000 orbits. In total these calculations took about 55 CPU days. 
\begin{figure}
    \centering
    \includegraphics[scale=0.28]{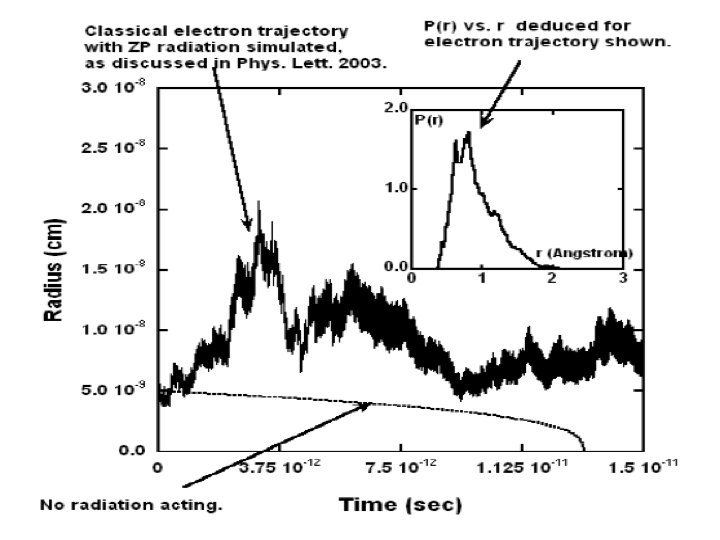}
    \caption{Typical plot of r vs time for one trajectory. The inset shows the probability density P(r) vs r for this trajectory. A lower curve of r vs. t shows that if no zero-point radiation was present, then atomic collapse would occur in about $1.3\times 10^{-11}$ sec.  The starting trajectories began at 0.53 \AA\ \cite{zhou and cole}.}
    \label{fig3}
\end{figure}

Fig. \ref{fig3} shows a typical trajectory as a function of time.  It shows fluctuations in the radius by a factor of about 3 for roughly 100,000 orbits.  As the inset shows, without the ZP field, the electron would have collapsed after about $1.3\times 10^{-11}$ sec, so clearly the ZP field is preventing collapse during the interval shown. 

\begin{figure}[h]
\centering 
\begin{subfigure}{}
\includegraphics[width=0.4\linewidth, height=5cm]{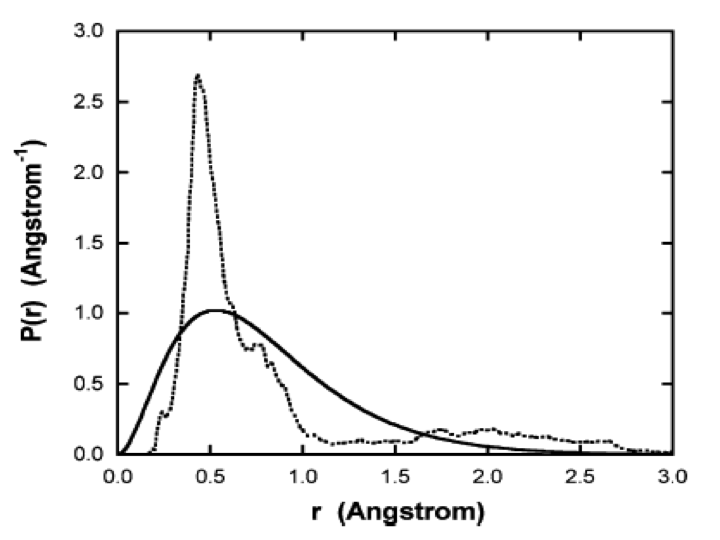} 
\label{fig:subim1}
\end{subfigure}
\begin{subfigure}{}
\includegraphics[width=0.4\linewidth, height=5cm]{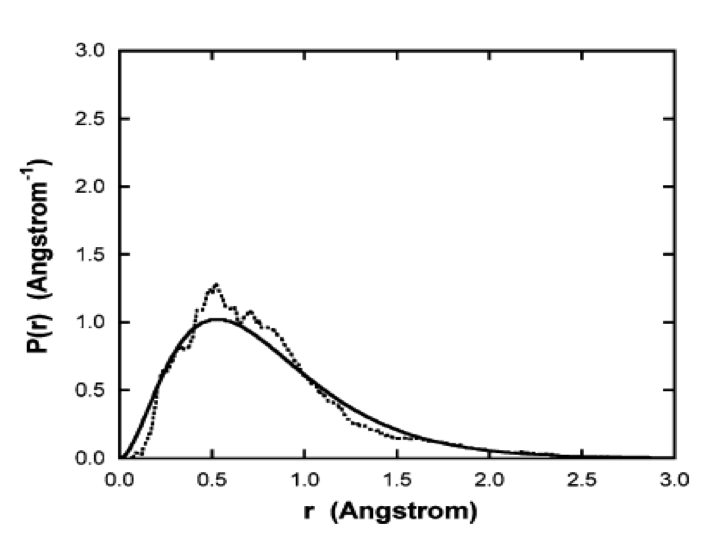}
\label{}
\end{subfigure}
 
\caption{Plots of the radial probability density vs. radius.  The solid line was calculated from the ground state Schrodinger wavefunction. The dotted curves are simulation results, with starting trajectories began at 0.53\AA\,  calculated as a time average for all 11 simulation runs from t=0 to the indicated time: left side, $1.417\times 10{^-12}$ sec, right side, $7.25 \times 10^{-12} sec$ \cite{zhou and cole}.}
\label{fig4}
\end{figure}

Fig. \ref{fig4} shows the radial probability distribution P(r) vs. radius, computed from a time average of the 11 simulations, up to the time indicated for each plot.  Also shown is the P(r) computed from the ground state Schrodinger wavefunction.  As time increases, the distribution approaches that predicted by quantum mechanics.

The simulations of Cole and Zou typically ran for about $10^{-11}$sec or roughly $100,000$ orbits and showed that as time increased, the radial distribution in $r$ tended to look more like that predicted by quantum mechanics. In subsequent simulations using 30 PCs, Cole increased the span of wavelengths used for the applied zero-point field from $10^{4}$ to $10^{6}$ and found ionization occurred unless they increased the precision of the computation \cite{cole v}. These preliminary results were only for about $10^{-12}$ sec, about one tenth the time of the previous work and showed general agreement with the previous work. Cole and Zou also explored fundamental interactions between circularly polarized (CP) electromagnetic radiation and the electron initially in orbit about the proton, demonstrating that a stable circular orbit could be obtained by judicious choice of the amplitude and phase of a single CP normally incident wave with the same frequency as the orbital motion, and that a stable elliptical orbit could be obtained with a superposition of CP waves consisting of many harmonics of the orbital motion\cite{colejsc}. Cole and Zou\cite{colesub} and Cole\cite{10} did an ensemble of simulations with different initial positions of the electrons, and explored resonances that arise for certain orbital conditions, deepening understanding about the details of SED simulations and the classical H atom, and possibly paving the way for understanding transitions in H atom energy levels with SED. 

The simulations by Nieuwenhuizen and Liska \cite{nie}, done over a decade after the pioneering work of Cole and Zou\cite{zhou and cole} used the vastly improved computing power ($5.6$ TFLOPS single precision floating point). They did a 3d simulation that included about $10^7$ plane waves. To simplify the computation they replaced the 3d sums over all $\bo{k}$ by 1d sums over frequency with Gaussian amplitudes chosen such that they reproduce the same correlation function in the limit where the frequency mesh vanishes $(N\rightarrow\infty)$. They also computed orbits for times that were twice as long, about $3\times 10^{-11}$ sec, with about $10^6$ orbits.

A 4th order Runge-Kutta algorithm was used with about 4000 iterations/orbit. The electric field was updated as a function of time ten times each orbit, with interpolation in between. Numerous methods were employed to check the computational accuracy of the results.

Fig. \ref{fig7} shows the energy in Bohr units and the eccentricity as a function of the time with units of $t_o=1/\omega_{0}$, the Bohr time $t_{0}=\hbar/Z^2 \alpha^{2}m c^2$, which is $(1/2\pi)$ of the QM orbital period. The energy of the ground state is $ \hbar \omega_{0}/2$. These simulations employed a moving cutoff equal to 2.5 times the orbital frequency. The energy varies from about 0.1 to -1.6 Bohr units (without radiative reaction the energy would be -.5). The eccentricity $\epsilon$ varies from about 0.05 to 0.95 (without radiative reaction $\epsilon$ would be 0 for a circular orbit).  

\begin{figure}[h]
\centering 
\begin{subfigure}{}
\includegraphics[width=0.4\linewidth, height=5cm]{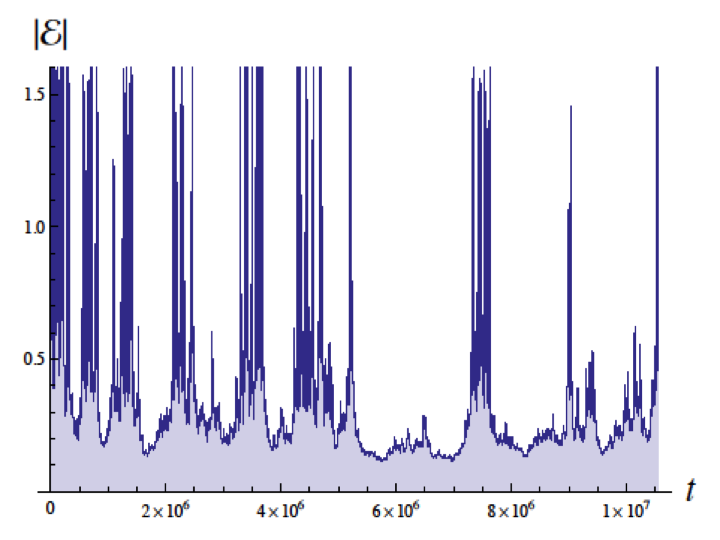} 
\label{fig:subim1}
\end{subfigure}
\begin{subfigure}{}
\includegraphics[width=0.4\linewidth, height=5cm]{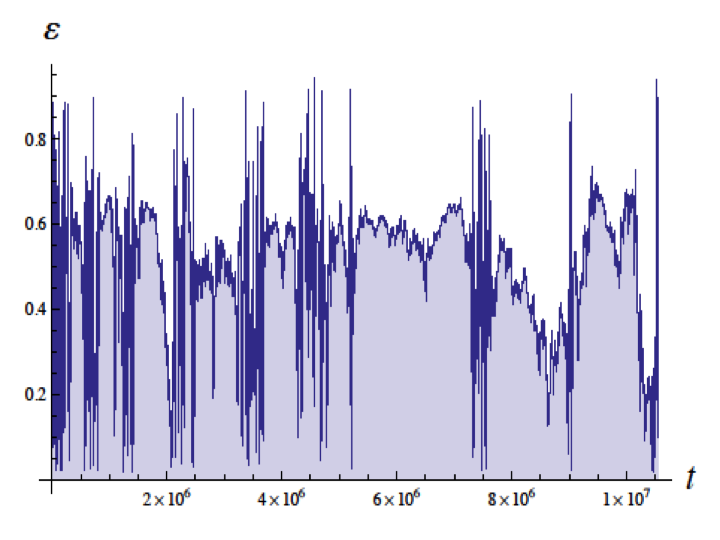}
\label{fig7}
\end{subfigure}
 
\caption{Plots of the energy E (left side) and eccentricity $\epsilon$ (right side) as a function of time in units of $t_0$ for Z=3, for $1.5 \times 10^{6}$ revolutions. The energy for the ground state in quantum mechanics would be 0.5 with the scale used (Bohr units)\cite{nie}.}
\label{fig7}
\end{figure}
\begin{figure}
\centering 
\begin{subfigure}{}
\includegraphics[width=0.4\linewidth, height=5cm]{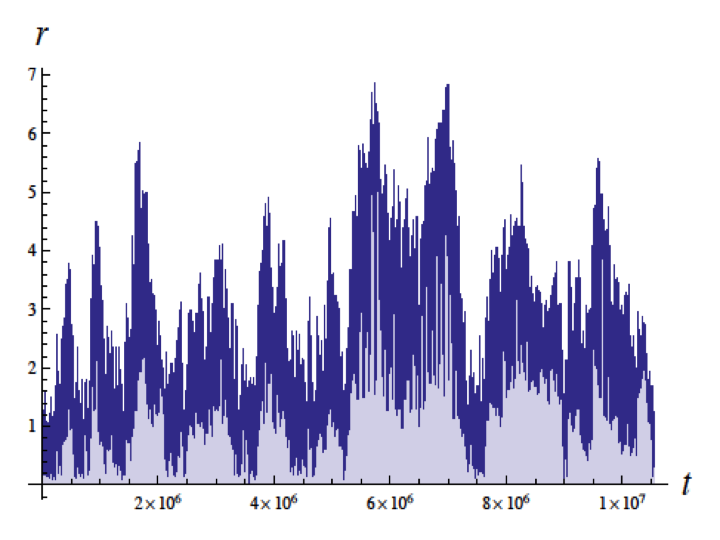} 
\label{fig:subim1}
\end{subfigure}
\begin{subfigure}{}
\includegraphics[width=0.4\linewidth, height=5cm]{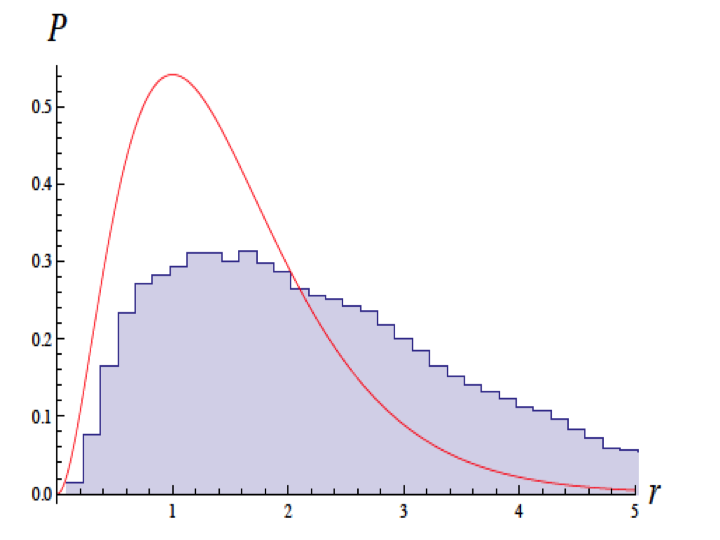}
\label{}
\end{subfigure}
 
\caption{Plots of the radius r of the electron orbit as a function of time in units of $t_0$ (left side) and a histogram of r (right side), for the same simulation as Fig. \ref{fig7}. The smooth curve is the probability distribution expected from quantum mechanics.\cite{nie}}
\label{fig9}
\end{figure}

Fig. \ref{fig9} shows the variation in the radius with time in units of $t_0$ and a corresponding histogram of the radius.  The radius varies very rapidly from 0.1 to 8 Bohr radii , while the eccentricity and energy remain somewhat more stable on longer timescales.  Unlike Cole and Zou \cite{zhou and cole}, they did not see a trend to reproduce the wavefunction, although their simulations suggested this might occur for shorter times scales.

\begin{figure}
\centering 
\begin{subfigure}{}
\includegraphics[width=0.4\linewidth, height=5cm]{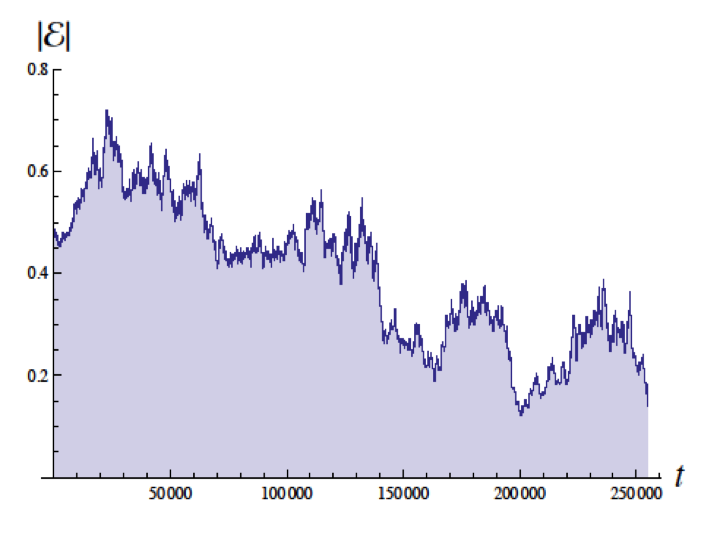} 
\label{fig:subim1}
\end{subfigure}
\begin{subfigure}{}
\includegraphics[width=0.4\linewidth, height=5cm]{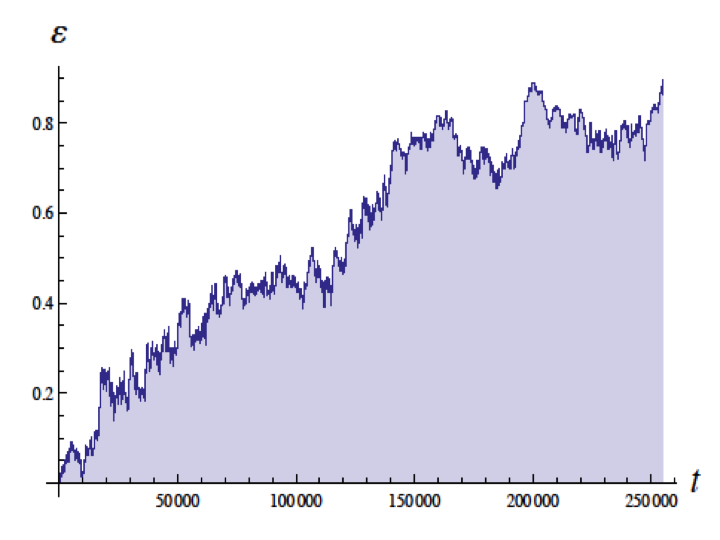}
\label{}
\end{subfigure}
 
\caption{Plots of the energy E (left side) and the eccentricity $\epsilon$ as a function of time for Z=1 with a fixed cutoff exposing the trend towards ionization at $E=0,\ \epsilon=1$.  The time window in 45 times shorter than in Fig. \ref{fig7} and Fig. \ref{fig9} \cite{nie}.}
\label{fig12}
\end{figure}

These plots suggest the possibility that the orbits may stabilize.  However, when the simulations are done for longer times of the order of $10^7t_0$, instabilities developed which eventually lead to ionization Fig.\ref{fig12}. When higher harmonics were included (4.5 and 6.5), ionization occurred earlier.  Ionization was defined as the moment when the electron stayed above $E= -0.05$ for at least $10^7t_0$.  The moment of ionization is not shown in the subsequent plots\cite{nie}. 

With an upgrade in computing power, allowing double precision and a fixed cutoff of the frequency spectrum of the random fields, ionization occurred at an even earlier time. The energy of the electron tended to decrease to zero while the eccentricity increased leading to ionization. It is not clear precisely why these computational upgrades led to ionization at significantly earlier times.

\begin{figure}
    \centering
    \includegraphics[scale=0.28]{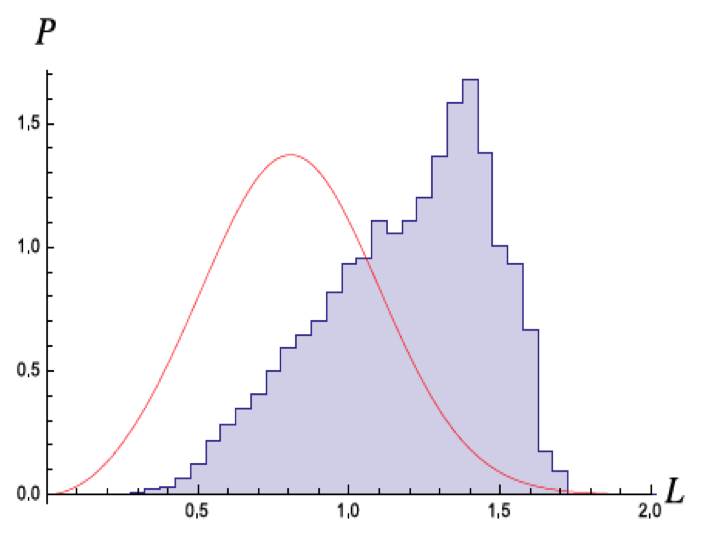}
    \caption{Distribution of the angular momentum L (in units of $\hbar$) from the
numerical simulation of the SED dynamics reported in \cite{nie}. Not much weight lies below
$L = 0.588$ , confirming that when such a value is reached at near-zero energy, self-ionisation
may occur rapidly and the run is ended. Full curve: the distribution of L from the conjecture
for the would-be stable ground state distribution from \cite{nieaip}.}
    \label{fig10}
\end{figure}

In a second set of simulations, Nieuwenhuizen and Liska included relativistic corrections and the effects of the magnetic fields\cite{nierel}.  Nevertheless, they obtained essentially the same results, ionization at early times. 
Niewenhuizen did a theoretical model of the H atom with the stochastic field, and computed the change in energy with each orbit of the electron\cite{nieent}. If the energy gets very low and the angular momenta goes below $0.588 \hbar$, then the orbits are very eccentric, and the electron gains energy with each rotation, leading to self-ionization.  In a simulation, as shown in Fig. \ref{fig10} they observed that the angular momentum seldom went below this critical value, in agreement with the predicted result.

\section{Discussion}
SED researchers have made significant progress in the very complex task of modeling the ground state of the H atom, and shown that for short times corresponding to tens of thousands of orbits the real stochastic zero-point field can provide the energy lost to radiation and maintain the orbital motion. However, as time passes and more regions of phase space are explored, the orbits tend to become highly elliptical, with lower angular momentum, and self-ionization occurs. The question arises: is this instability due: 1) to the specific implementations of SED, including the approximations and numerical methods used, or 2) is this instability inherent in the SED approach to the H atom, or 3) is it more fundamental in nature, and perhaps reflects a characteristic chaos arising from non-linear potentials \cite{cole fop}\cite{10}, as observed, for example, in comparisons of classical and quantum models of photo-dissociation with the Morse potential for diatomic molecules \cite{milonnimorse}.  For a variety of non-linear systems there are regions of classical phase-space characterized by chaos. 

Contrasting the approach of quantum mechanics and quantum electrodynamics to that of SED is illuminating. In QED, the various physical phenomena, including the zero-point field, the fundamental properties of the electron, the presence of radiation, the properties of the Coulomb field, and the stability of atoms, are all parsed in such a way that the system is mathematically tractable and well behaved solutions, reflecting the symmetry of the 1/r potential, are found. On the other hand it appears that the current SED approach lumps all these phenomena together in the Abraham-Lorentz equation with the zero-point field, which leads to very challenging mathematical entanglements that mask the symmetry of the solutions. For example, in QED, the self interaction leads to a big effect on the electron: mass renormalization.  The remaining radiative shift for the atom is very small.  In SED, the radiative effects in the Abraham-Lorentz equation lead to weak but persistent forces, whereas the vacuum field forces are much larger and fluctuate greatly from positive to negative values, making it difficult to obtain a stable orbit.

There are many more issues that SED needs to address beyond the attainment of a stable orbit. Can SED develop a set of states for the H atom that are isomorphic to those predicted by QM, with characteristic quantized values of angular momentum and energy? with the same degree of degeneracy? Can SED predict the transitions between energy levels? Can SED predict the results of measurements on the atom? It is very difficult to see a path for attaining these objectives with the current SED approach.  For a first step, for example, SED needs not only to obtain stable ground state orbit, but, to probably predict an orbit that is similar to the quantum ground state which has a spherically symmetric probability distribution with quantized angular momentum of zero. Perhaps this might be obtained with a superposition of suitable initial conditions.  But considering superpositions leads to ideas characteristic of quantum mechanics. This raises the question: does making SED predict the H atom correctly require the modification of SED so that it becomes essentially a reformulation of quantum mechanics? Perhaps SED, which predicts particles with definite orbits, has a relationship to Bohmian mechanics, which also describes definite orbits, has a non-local potential, and predicts measurements that agree with those of quantum mechanics\cite{holland}. 
\vspace{1pt}
\acknowledgments{The author would like to thank Peter Milonni for his encouragement and his comments on this paper, and to thank Dan Cole for sharing references and research I was unaware of, and for his extraordinarily careful reading and numerous insightful suggestions and comments.}



\end{document}